\documentstyle[prb,aps,preprint]{revtex}
\begin{document}
\draft
\title{Quantum states and specific heat of low-density He gas adsorbed
within the carbon nanotube interstitial channels: Band structure effects
and potential dependence}

\author{Antonio \v{S}iber$^1$\thanks{Corresponding
author. E-mail: asiber@ifs.hr} and Hrvoje Buljan$^2$}
\address{$^1$ Institute of Physics, P.O. Box 304, 10001
Zagreb, Croatia \\
$^2$ Department of Physics, Faculty of Science, University of Zagreb, PP
332, 10001 Zagreb, Croatia
}
\maketitle

\begin{abstract}
We calculate the energy-band structure of a He atom trapped within
the interstitial channel between close-packed nanotubes within a bundle 
and its influence on the specific heat of the adsorbed gas. A robust
prediction of our calculations is that the contribution of the low-density
adsorbed gas to the specific heat of the nanotube material shows
pronounced nonmonotonic variations with temperature. These variations are
shown to be closely related to the band gaps in the adsorbate density of
states.
\end{abstract}


\begin{center}
1. INTRODUCTION \\
\end{center}

The adsorption of gases in and on the bundles of carbon nanotubes has
recently attracted much attention
\cite{ColeCol,Boni1,ColePRL,JLow,uptake,cylind,Teizer}. The carbon
nanotube materials provide an excellent opportunity for a study of
adsorption \cite{Bruchbook} in confined geometries. This phenomenon is not
only of academic interest - a motivation for its study is potential
application of the carbon nanotube materials
for a very efficient storage of gases \cite{ColeCol,JLow,uptake,Teizer}.

$^4$He atoms adsorbed in strongly bound positions have been 
detected by measurements of desorption from single
wall nanotube (SWNT) bundles \cite{Teizer}.
Recent low temperature
specific heat measurements on SWNT bundles
\cite{JHone,Las1} have detected a contribution to the specific heat
which is very sensitive to the amount of
adsorbed gases, in particular $^4$He. Motivated by these experimental
results, here we theoretically analyse the behavior of He atoms adsorbed
in the samples consisting of SWNT bundles. In particular, we
investigate the quantum states available
for the adsorption of He atoms in the interstitial channels between the
SWNTs, close-packed in a triangular lattice. We analyse the
consequences of specificities of energy spectrum of adsorbed He atoms on
the specific heat pertaining to the adsorbed He gas.

The effects of interactions between interstitial He adsorbates and their
consequences
on the specific heat have been examined in
Ref. \onlinecite{ColePRL}. Transition
of He gas adsorbed in the interstitial channels to an anisotropic
condensed phase has been predicted as a result of attraction between
atoms in neighboring channels. 
In this work we concentrate on the interaction of He atoms with a
surrounding nanotube medium and neglect He-He interactions.

The main results and the outline of the article are as follows. In
Sec. 2, we propose a model potential for the interaction of He atoms
with surrounding medium. Our model of the potential confining He adsorbate
to an interstitial channel is sufficiently refined
to allow the investigation of the influence of details of He confinement
on the specific heat of adsorbate gas. We shall demonstrate that the
periodicity of the potential along the channel axis causes gaps in the
adsorbate density of states. These gaps increase with the magnitude
of corrugation of the interaction potential along the interstitial
channel axis. As a consequence, in Sec. 3
we shall show that the specific heat pertaining to the
adsorbate gas shows nonmonotonic behavior
with temperature. The precise dependence of specific heat on the
temperature will be shown to be determined by the corrugation of the
interaction potential.

Our approach is similar in spirit to the one presented in
Ref. \onlinecite{Colegraph}, where the specific heat of low-density He
films on
graphite has been examined. The most important results of this article are
summarized in Sec. 4.

\begin{center}
2. ADSORPTION POTENTIAL AND BAND STRUCTURE CALCULATION \\
\end{center}

The Schr\"{o}dinger equation for the wavefunction $\Psi_{\kappa}
(x,y,z)$ of a single adsorbate within the interstitial channel is
\begin{equation}
\left [ -\frac{\hbar ^2}{2M} \nabla ^2 + V(x,y,z) \right ] \Psi_{\kappa}
(x,y,z) = E_{\kappa} \Psi_{\kappa} (x,y,z),
\label{eq:Schsimple}
\end{equation}
where $M$ is the mass of adsorbate and $\kappa$ denotes a set of
quantum numbers needed for the specification of the quantum state of the
adsorbate.

To find the quantum states supported by equation
(\ref{eq:Schsimple}) one needs to construct the potential
$V(x,y,z)$. However, the construction of the relevant potential is
somewhat problematic for two reasons. First, the potential describing the
interaction of He atom with the carbon nanotube in not known with great
precision and
often the model potentials based on the assumption of pairwise additivity
of binary He-C interactions are used \cite{JLow,uptake}. Second, the
details of arrangement of nanotubes within a bundle are also not known. In
particular, the relative axial alignment of nanotubes (relative
axial offsets) within a bundle is
not known \cite{Boni1}. As shown in Ref. \onlinecite{Boni1}, different
nanotube
alignments can produce very different potentials for adsorption of He atom
in the interstitial channels. It is also likely that bundles contain a
mixture of tubes with different wrapping angles
\cite{wrap,NaturLas}. Details of this mixture
may vary depending on a particular bundle in the sample. Furthermore, the
distribution of nanotube diameters in the sample has a finite width
\cite{Rols1,Thess}. All these facts
imply that a detailed information on the potential is at present not
available and that the relevant potential depends on the
interstitial channel in question.

Here we propose a model potential $V(x,y,z)$ which captures all
the essential physics involved, yet simple enough to allow an
efficient and transparent calculation of the quantum states and predict
the dependence of specific heat on temperature. According to references
\onlinecite{JLow} and \onlinecite{uptake}, the effective potential
confining the motion of
He atom in the plane perpendicular to the channel axis can be represented
by a paraboloid-like shape with the minimum at the geometrical center of
the channel (see Fig. 1 of Ref. \onlinecite{JLow}). The potential varies
along the channel axis, reflecting the variations of the nanotube
electronic density in this direction \cite{Bruchbook}. The simplest
potential which displays the described behavior is
\begin{eqnarray}
V(x,y,z) &=& D + \alpha (x^2 + y^2) + 2 V_g (x,y) \cos (gz) \nonumber \\
         &=& V_0 (x,y) + 2 V_g (x,y) \cos (gz),
\end{eqnarray}
where $g = 2 \pi /a$ is the inverse lattice vector associated with the
periodicity of the potential along the channel axis ($z$ direction) and
the factor of 2 in front of $V_g(x,y)$ term has been introduced for
convenience. The origin of $x$ and $y$ axes is positioned at the center of
the channel and the channel axis is perpendicular to the $xy$ plane. $D$
defines the minimum of the potential at the center of the channel, in
absence of the corrugation term, $V_g(x,y)$. Equation
(\ref{eq:Schsimple}) can be solved by writing the wavefunction as
\begin{equation}
\Psi_{\lambda}^k (x,y,z) = \sum _{m,n,G} \beta _{m,n,k+G}^{\lambda}
\Phi_{m,n,k+G}(x,y,z),
\label{eq:psi_ans}
\end{equation}
where 
\begin{eqnarray}
\Phi_{m,n,k+G} (x,y,z) &=& \frac {\delta} {\sqrt{\pi 2^{m+n} m! n! L_z}}
H_m(\delta
x) H_n (\delta y) \exp \left( -\frac{1}{2} \delta^2(x^2+y^2) \right) e
^{i(k+G)z} \nonumber \\
&=& |m,n;k+G>
\end{eqnarray}
is the solution of the hamiltonian containing only $V_0 (x,y)$ part of
the potential. Here, $L_z$ is the box quantization
length along the axis of the channel, $H_{m,n}$ are the Hermite
polinomials, $\delta ^4  = 2 M \alpha / \hbar ^2$, $m$ and $n$ are the
harmonic oscillator quantum numbers, $k$ is the wavevector associated with
the motion of He atoms in $z$ direction, and $G$ is a reciprocal lattice
vector given as $G=p g$, with $p$ integer, associated with the
periodicity of the potential in $z$ direction. The Schr\"{o}dinger
equation (\ref{eq:Schsimple}) reduces to
matrix equation
\begin{eqnarray}
& &\sum_{m',n',G'} \left [ E_{m',n'} + \frac{\hbar^2}{2M} (k+G')^2 -
E_{\lambda} \right ] \delta_{m,m'} \delta_{n,n'} \delta_{G,G'}
\beta_{m,n,k+G}^{\lambda} \nonumber \\
&+& \sum_{m',n',G'} <m,n;k+G|2 V_g(x,y) \cos(gz)|m',n';k+G'>
\beta_{m,n,k+G}^{\lambda} = 0.
\label{eq:matrix}
\end{eqnarray}
$E_{m',n'}$ are the eigen energies of the hamiltonian without the
corrugation part of the potential and are given as
\begin{equation}
E_{m',n'} = D + \hbar \omega_0 (m'+n'+1),
\label{eq:harmo}
\end{equation}
where $\omega_0=\sqrt{2 \alpha / M}$ is the characteristic frequency of
the harmonic oscillator.
The spectrum of energies of adsorbed He atom can be found by looking for
nontrivial solutions of equation (\ref{eq:matrix}). For fixed $k$, this
produces a number of different solutions (bands) denoted by index
$\lambda$.

By systematic examination of a range of $k$ values, it is
possible to obtain the full band structure, $E_{\lambda}(k)$. Similar
approach has been used for a calculation of band structure of He atoms
adsorbed on the surface of graphite \cite{Colegraph} and NaCl
\cite{Vargas}.

We parametrise the corrugation term in the potential as
\begin{equation}
V_g(x,y) = c_0 + c_2 (x^2+y^2).
\end{equation}
The matrix elements of the corrugation term can be cast in the analytic
form as
\begin{eqnarray}
V_{m,n,G} ^{m',n',G'} &=& <m,n;k+G|2 V_g(x,y) \cos(gz)|m',n';k+G'>
\nonumber \\ 
&=&\delta _{G,G' \pm g} \left \{
\left[ c_0 + \frac{c_2}{\delta^2}(m'+n'+1) \right] \delta_{m,m'}
\delta_{n,n'} \right . \nonumber \\
&+& \frac{c_2}{2 \delta^2} \left[ \delta_{n,n'} \left (
\sqrt{(m'+1)(m'+2)} \delta _{m,m'+2} + \sqrt{m'(m'-1)} \delta_{m,m'-2}
\right ) \right. \nonumber \\
&+& \left. \left. \delta _{m,m'} \left(\sqrt{(n'+1)(n'+2)} \delta_{n,n'+2}
+ \sqrt{n'(n'-1)} \delta_{n,n'-2} \right ) \right ] \right \}.
\end{eqnarray}
The parameters of the proposed model of interaction potential can be fixed
by examining the
potential obtained as a pairwise sum of binary He-C site interactions. We
model
the effective He-C site interaction with a Lennard-Jones (LJ) form,
4$\epsilon [(\sigma /r)^{12} - (\sigma /r)^6]$, where $r=|{\bf r}|$
denotes the distance
between He atom and a C site.
Various different forms of effective He-C
potential, based on the analysis of He-graphite interaction, have been
proposed in the literature \cite{ColeFrankl,Toigo}. The most important
deficiency of the LJ form is the assumption that the potential depends
only on the magnitude of ${\bf r}$ \cite{ColeFrankl}. However, in the view
of a lack of knowledge on the precise geometry of a bundle, the LJ form
should serve well in obtaining the main features of the interaction
potential. The LJ parameters we adopt are $\epsilon =1.34$ meV,
$\sigma=2.75$ \AA. This choice of parameters was suggested in
Ref. \onlinecite{Garcia}, where the LJ form of the potential has been
optimized to reproduce the experimental data on $^4$He scattering from the
basal (1000) plane of graphite. These parameters are slightly different
from the ones reported in Ref. \onlinecite{uptake} ($\epsilon =1.46$ meV,
$\sigma=2.98$ \AA) which were obtained from the semiempirical combining
rules \cite{Bruchbook}.

In the following, we consider (10,10), "armchair" carbon
nanotubes \cite{wrap} with a diameter of $d=$ 13.8 \AA. The centers of
the nanotubes are separated by 13.8 \AA +3.2 \AA = 17.0 \AA, where the
value of the intertube separation of 3.2 \AA \hspace{0.7mm} has been
adopted, in agreement with experimental findings
\cite{Thess} and also with Refs. \onlinecite{JLow} and
\onlinecite{uptake}. Assuming that all the tubes surrounding a channel are
axially
aligned, and that the whole structure of the bundle can be obtained from
a single tube by applying a two dimensional translation characterised by a
general two dimensional lattice vector given as ${\bf t} = i {\bf t}_1 + j
{\bf t}_2$, where $i$ and $j$ are integers and ${\bf t}_1$ and ${\bf t}_2$
are basis vectors of a triangular lattice of the bundle, the parameters
consistent with our model of the potential are $D=-36.8$ meV, $\alpha =
44.7$ meV/\AA$^2$, $c_0 = - 0.05$ meV, $c_2 = 10.7$ meV/\AA$^2$ and $g=2
\pi / 2.5$ \AA $=2.51$ 1/\AA. This set of parameters provides an
excellent fit to the angular average of the total potential \cite{JLow}.

The density of states per unit length of the channel can be obtained from 
\begin{equation}
g(E) = \frac{1}{L_z}\sum_{\lambda, k} \delta (E - E_{\lambda}(k)).
\label{eq:dens}
\end{equation}
Some care should be taken in evaluation of Eq. (\ref{eq:dens}) to
properly account for the degeneracy present in the energy spectrum,
$E_{\lambda}(k)$. In Fig. \ref{fig:fig1} we present the band structure and
the corresponding density of states obtained from the determined set of
potential parameters. For convenience of presentation, the delta functions
in equation (\ref{eq:dens}) have been
broadened to gaussians of width 0.03 meV. Note the enhancement of
the lowest band effective mass of He atom, $M^* = 2.37 M = 9.5$ amu and
the band gap between -27.75 meV and -25.5 meV. The bandwidth of the lowest
energy band is 0.19 meV (2.2 K). This should be compared with the value of
0.18 K found for (18,0) tubes in Ref. \onlinecite{ColePRL}. The difference
between these two values is determined in part by the different
periodicities (cell lengths) of the "zigzag" and "armchair" carbon
nanotubes.

The mass enhancement should be
compared with the one found by authors of Ref. \onlinecite{Boni1},  $M^* =
1.3 M$. The difference between their and our value can be explained by
somewhat larger intertube
separation presumed in the calculations in Ref. \onlinecite{Boni1} (3.3
\AA, whereas we used 3.2 \AA). As shown in
Ref. \onlinecite{Boni1} this parameter strongly influences the magnitude
of the potential corrugation and thus also the effective mass
enhancement. The ground-state energy, $E_{\Gamma}$, of adsorbed $^4$He
atoms as found from our model potential is -27.94 meV. The
ground-state energy found in references \onlinecite{JLow} and
\onlinecite{uptake} for nanotubes of the same diameter (13.8 \AA), but for
the interaction potential which does not exhibit corrugation along the
channel axis was -29.07 meV and -33.3 meV, respectively. This shows 
consistence of our potential with the potentials
previously proposed in the literature.
The experimental estimate of the binding energy based on desorption
measurements \cite{Teizer} was reported to be -28.5 meV, in good agreement
with the value we obtain. The subsequent reevaluation \cite{Teizer2} of
the experimental data first presented in Ref. \onlinecite{Teizer} yielded
a smaller binding energy (-19.8 meV). The authors of
Ref. \onlinecite{Hallock} suggested
that this value of binding energy may be compatible with adsorption of He
on the outer grooves of
the bundle. Our results should be of help to determine the relevant
adsorption sites.

One of the attractive features of our potential model is that
the various different arrangements of nanotubes surrounding a channel can
be easily (but approximately) simulated by changing the parameters of the
potential, especially those representing the corrugation of 
the potential. Additionally, the effects of possible inadequacies of the
binary He-C site potential (such as the lack of anisotropy in the
effective interaction) can be avoided by treating the parameters $c_0$
and $c_2$ of the potential as variables. In Fig.
\ref{fig:fig2} we present the behavior of the density of states as a
function of parameter $c_2$ which defines the magnitude of the potential
corrugation. As expected, for smoother potentials,
the band gaps, binding energy and effective mass are smaller. In
particular, the lowest band mass enhancements are 2.37, 1.23 and 1.02 for
the calculations presented in panels a), b) and c) of Fig. \ref{fig:fig2},
respectively.

\begin{center}
3. SPECIFIC HEAT OF LOW DENSITY $^4$He INTERSTITIAL GAS\\
\end{center}

Knowing the density of states, we can calculate the
isosteric specific heat of He atoms adsorbed in the nanotube
interstitial channels. The approach we pursue is similar to the
calculation of specific heat of the adsorbed (sub)monolayer of
noninteracting atoms on a corrugated surface discussed in Ref.
\onlinecite{Colegraph}. The influence of adsorbate-adsorbate interactions
on the specific heat of the He overlayer on graphite has been studied in
Ref. \onlinecite{Siddon}.

The total energy of a system of N noninteracting
$^4$He atoms is given by
\begin{equation}
U = L_z \int_{E_{\Gamma}}^{\infty} g(E) E f(E,T) dE.
\end{equation}
The Bose-Einstein distribution function in the above equation is given by
\begin{equation}
f(E,T) = \frac{1}{\exp \left( \frac{E-\mu}{k_B T} \right) -1 },
\end{equation}
where $k_B$ is the Boltzmann constant and $T$ is temperature.
The chemical potential $\mu$ can be determined from the
requirement of the conservation of number of adsorbates, $N$,
\begin{equation}
N = L_z \int_{E_{\Gamma}}^{\infty} g(E) f(E,T) dE.
\label{eq:Neq}
\end{equation}
Solution of Eq. (\ref{eq:Neq}) results in the temperature dependence of
chemical potential $\mu$ for a fixed total number of $^4$He atoms.
The isosteric specific heat $C/N$ can be obtained from
\begin{equation}
\frac{C}{N} = \frac{1}{N}  \left ( \frac{dU}{dT} \right )_{N}
= L_z
\frac{k_B}{(k_B T)^2} \frac{I_2 - I_1^2/I_0}{N},
\label{eq:ccalc}
\end{equation}
where integral quantities $I_j$ are defined as
\begin{equation}
I_j = \int_{E_{\Gamma}}^{\infty} g(E) E^j \exp \left[ \frac{E-\mu
(T)}{k_BT}
\right ] f^2(E,T) dE, \ j=0,1,2,
\label{eq:is}
\end{equation}
in accordance with the notation of Ref. \onlinecite{Colegraph}. Equations
(\ref{eq:ccalc}) and (\ref{eq:is}) are very convenient for numerical
implementation once the dependence of chemical potential on temperature,
$\mu(T)$ has been found from Eq. (\ref{eq:Neq}).

In Fig. \ref{fig:fig3} we plot the isosteric specific heat as a function
of temperature for three different average adsorbate linear densities and
with potential parameters obtained from the assumption of perfect
alignment of the tubes. The dashed (full) line corresponds to average
interadsorbate distance of 1 nm (2 nm). It is interesting to note here
that the mean He-He distance along the channel estimated from experimental
data in Ref. \onlinecite{Teizer} was found to be about 3 nm. The
dash-dotted line in Fig. \ref{fig:fig3} corresponds to this average linear
$^4$He density.

As noted in Ref. \onlinecite{Boni1}, perfectly aligned tubes result in the
highest corrugation of the potential seen by He adsorbate. In this
respect, it is interesting to see how the specific heat changes as a
function of a potential corrugation along the channel axis. 
In Fig. \ref{fig:fig4} we
plot the variation of specific heat with temperature for three different
corrugation parameters $c_2$ and we fix the average linear density of He
adsorbates to $N/L_z = 0.1$ 1/\AA. All three curves display a nonmonotonic
behavior and a minimum in specific heat around $T=4$ K is observed. At
higher temperatures all curves approach to the value of 0.5 (thin dotted
line), reflecting a specific heat characteristic of a particle with one
translational degree of freedom. For even higher temperatures ($\sim$
20 K, not shown in the figure) the
excitation of higher harmonic oscillator levels [$m'$ or $n'$ in
Eq. (\ref{eq:harmo}) different from zero] becomes possible and the
specific heat becomes larger than 0.5. For such high temperatures,
desorption of He from the sample becomes probable \cite{Teizer,JHone}. The
minimum in specific heat occurs due to the existence of the band gap in
the density of states \cite{Colegraph,Bruchbook}.

As the band
gap becomes smaller (i.e. for lower corrugations of the potential) the
minimum in specific heat becomes less pronounced and practically disappears
for the smallest corrugation of the potential considered in the
calculations presented in Fig. \ref{fig:fig4}. It is interesting to note
that our results
show similar trends as the ones presented in
Ref. \onlinecite{Colegraph}. Due to
the different dimensionality in our case, the specific heat is a factor of
about 2 smaller than the one calculated for rare overlayer of He atoms on
graphite \cite{Colegraph}. Also, due to the effective magnification of the
corrugation in a restricted geometry of the interstitial channel, the
variation of the specific heat with temperature is significantly more
pronounced.

\begin{center}
4. SUMMARY AND CONCLUSIONS \\
\end{center}

A robust prediction of our model is appearance of band gaps in the density
of states of He adsorbates in the interstitial channels. This prediction
is based on the assumption of periodicity of the potential along the
channel axis and it holds irrespective of the precise alignment of the
tubes surrounding a channel, as long as all three tubes have the same
internal symmetry. 
The existence of the band gap causes a nonmonotonic
behavior of the specific heat associated with the interstitial He gas, an
effect previously observed for low-density overlayer of He atoms on a
graphite surface \cite{Colegraph}. We observe an appearance of the minimum
in specific heat around $T=4$ K. As the corrugation of the
potential along the channel axis smoothens, the minimum in the
specific heat becomes less dramatic and ultimately vanishes. The effects
we predict should be easily tested experimentally in the regimes where the
specific heat of the adsorbates can be easily distinguished from the
overall specific heat (for low temperatures and not too low concentrations
of He adsorbates \cite{JHone,Las1}).

\begin{center}
ACKNOWLEDGEMENTS\\
\end{center}

A. \v{S}. acknowledges stimulating discussions with K. Biljakovi\'{c} and
J.C. Lasjaunias.



\begin{figure}
\caption{
a) Calculated band structure of a $^4$He atom adsorbed in the
interstitial channel. b) The corresponding density of states per unit
length.
}
\label{fig:fig1}
\end{figure}

\begin{figure}
\caption{Density of states of $^4$He atoms adsorbed in the interstitial
channel as a function of the magnitude of corrugation of interaction
potential ($c_2$ parameter). a) $c_0=-0.05$ meV, $c_2$=10.7
meV/\AA$^2$. b) $c_0=-0.05$ meV, $c_2$=5 meV/\AA$^2$. c) $c_0=-0.05$ meV,
$c_2$=2 meV/\AA$^2$.
} 
\label{fig:fig2}
\end{figure}

\begin{figure}
\caption{Isosteric specific heat of $^4$He gas within the nanotube 
interstitial channels as a function of the sample temperature for three
different average He linear densities, $N/L_z$. Dash-dotted
line: $N/L_z$=0.033 1/\AA. Full line: $N/L_z$=0.05 1/\AA. Dashed
line: $N/L_z$=0.1 1/\AA.
} 
\label{fig:fig3}
\end{figure}

\begin{figure}
\caption{Dependence of specific heat of $^4$ He gas on the corrugation of
the interaction potential along the channel axis. Full line: $c_0=-0.05$
meV, $c_2$=10.7 meV/\AA$^2$. Dashed line: $c_0=-0.05$ meV, $c_2$=5
meV/\AA$^2$. Dash-dotted line: $c_0=-0.05$ meV,
$c_2$=2 meV/\AA$^2$. In these calculations, average linear density of
$^4$He
atoms was set to 0.1 1/\AA.} 
\label{fig:fig4}
\end{figure}

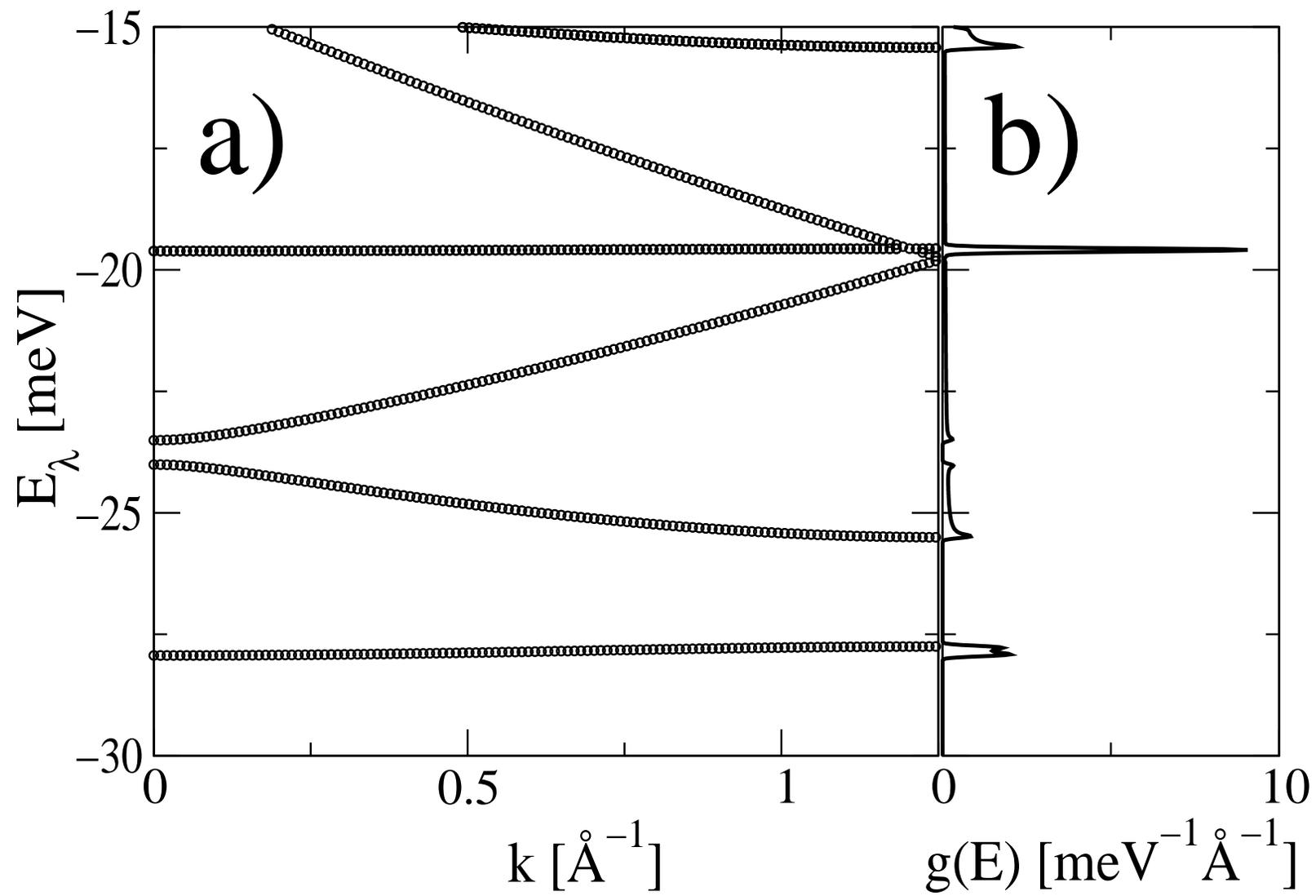
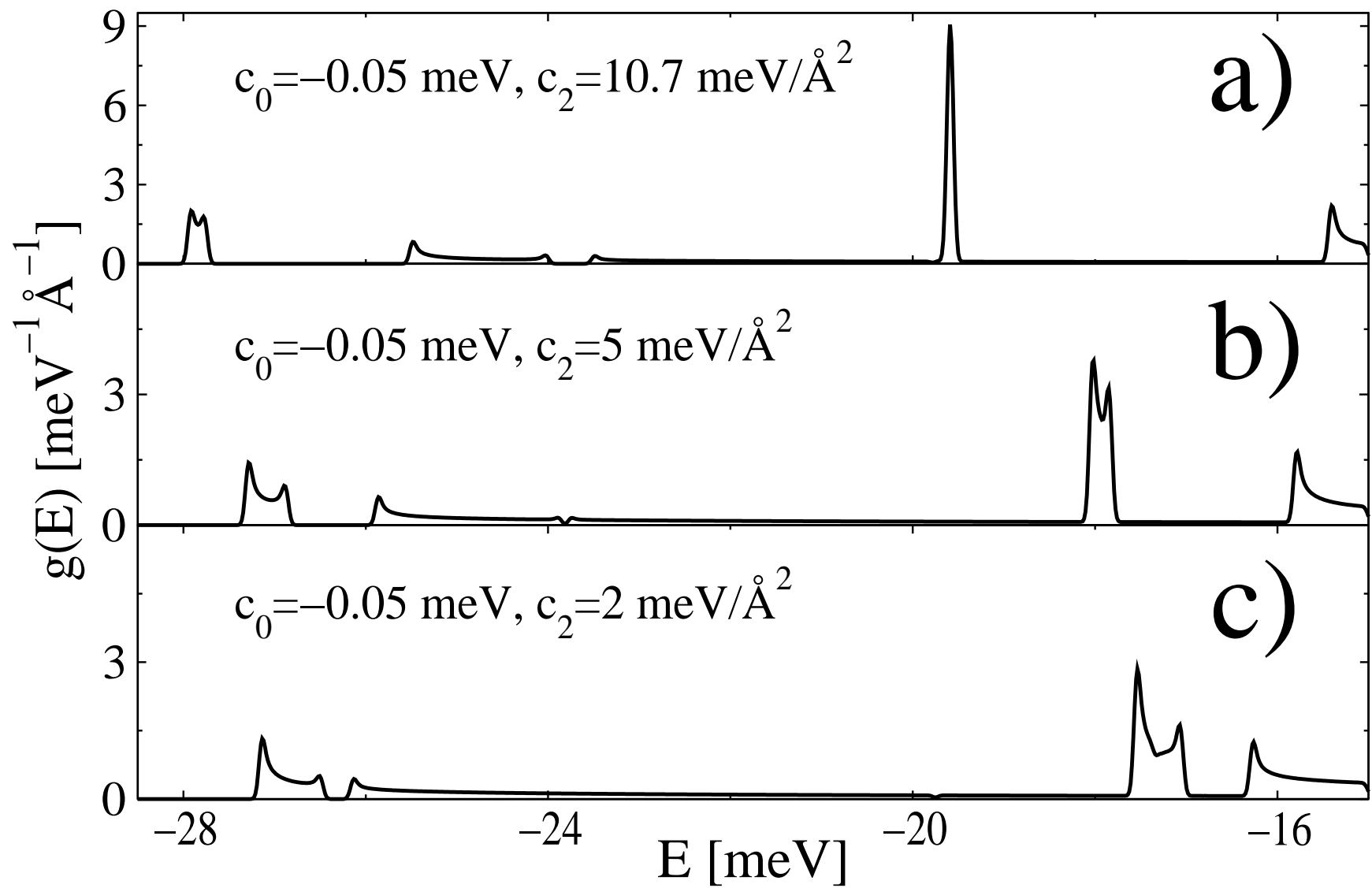
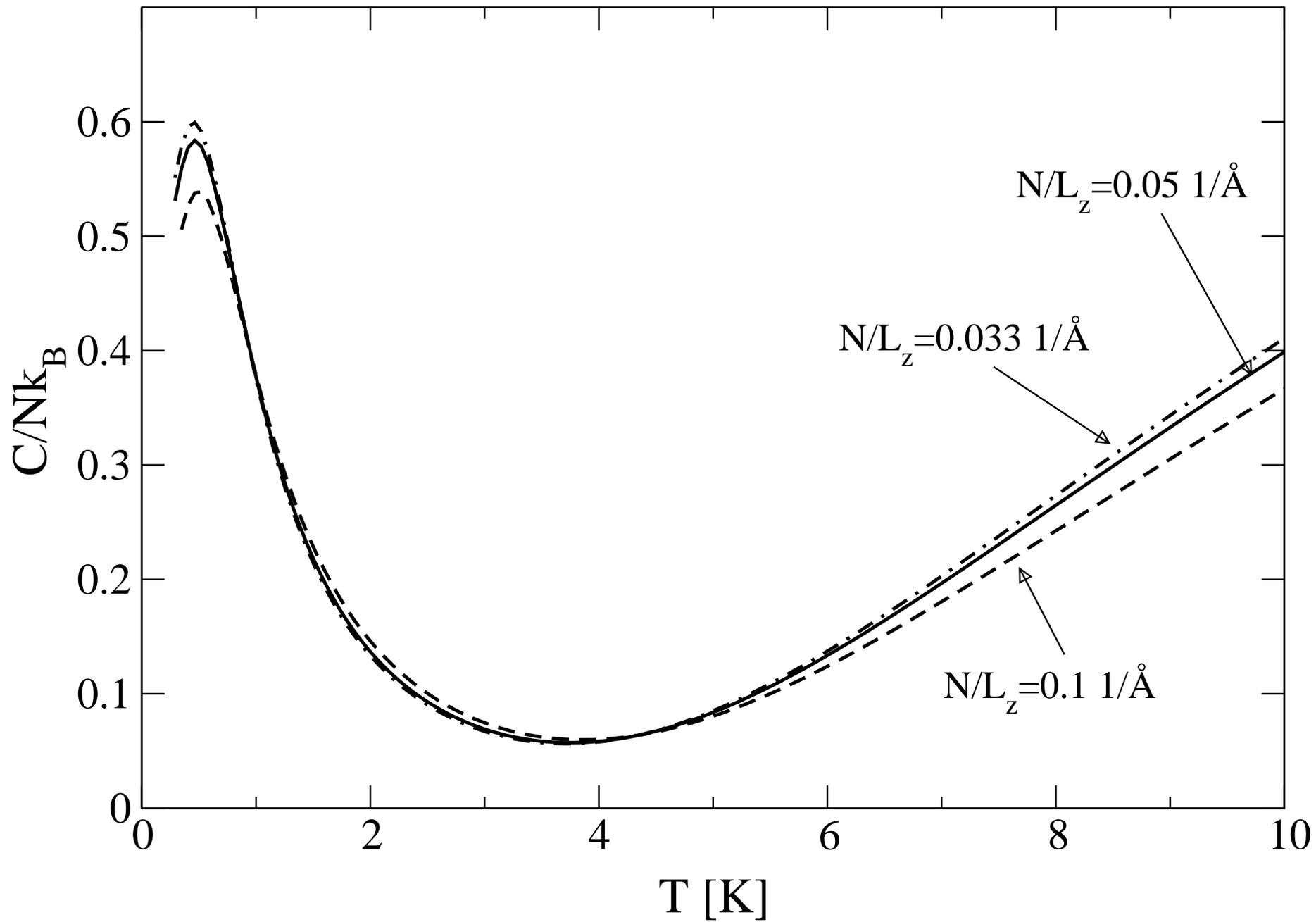
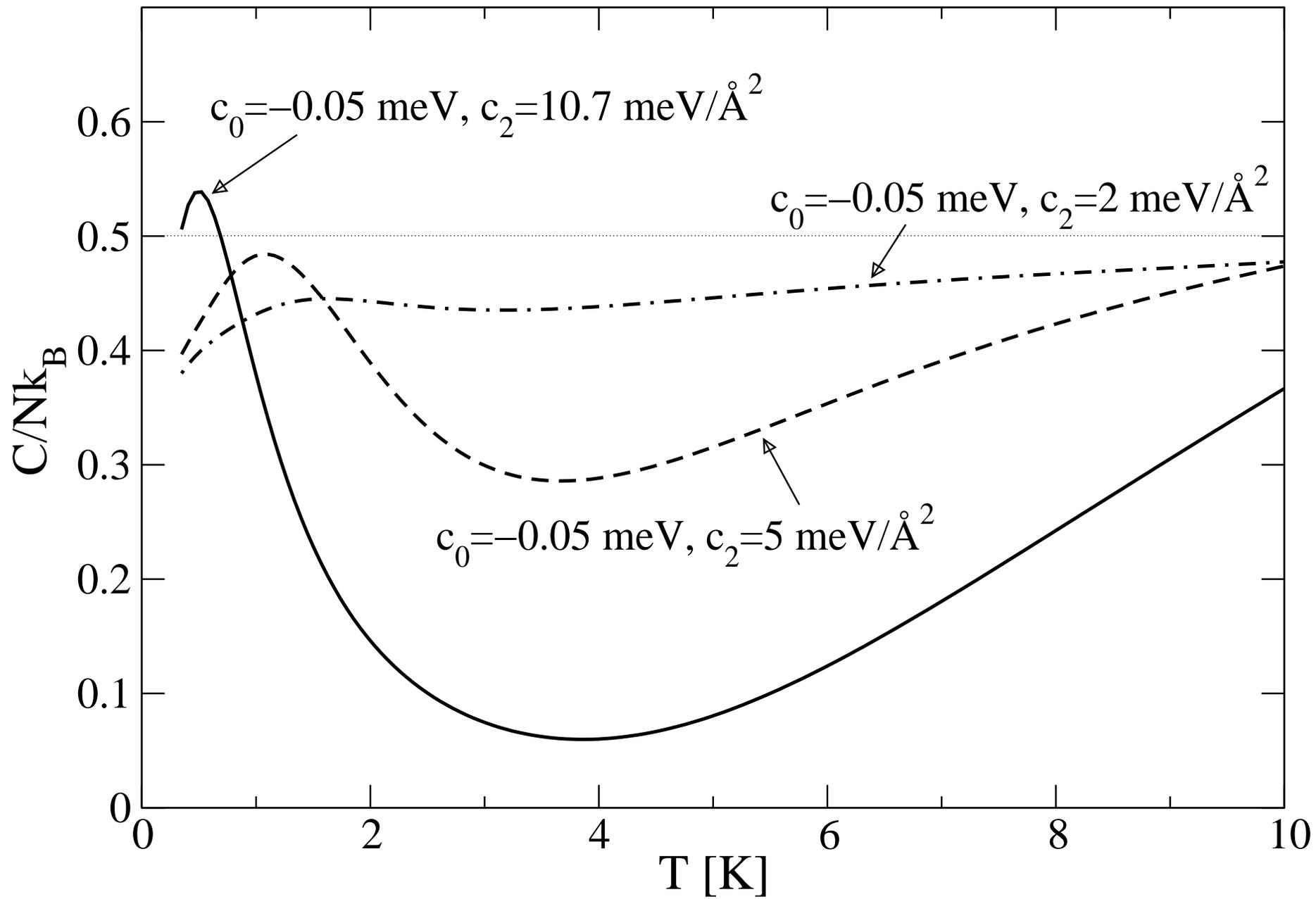
\end{document}